\documentclass[11pt,twoside]{article}
\usepackage{asp2014}

\aspSuppressVolSlug
\resetcounters

\begin{document}

\title{RCSEDv2: Open-source web tools for visualization of imaging and spectral data}

\author{Vladislav Klochkov$^{1,2}$, Ivan Katkov$^{1,3}$, Igor Chilingarian$^{1,4}$, Kirill Grishin$^{5,1}$, Anastasia Kasparova$^1$, Vladimir Goradzhanov$^{1,2}$, Victoria Toptun$^{1,2}$, Evgenii Rubtsov$^{1,2}$ and Sviatoslav Borisov$^{1,6}$}
\affil{$^1$Sternberg Astronomical Institute, Moscow State University, Moscow, Russia }
\affil{$^2$Department of Physics, Moscow State University, Moscow, Russia }
\affil{$^3$New York University Abu Dhabi, Abu Dhabi, UAE}
\affil{$^4$Center for Astrophysics -- Harvard and Smithsonian, Cambridge, MA, USA}
\affil{$^5$Universit\'e de Paris, CNRS, Astroparticule et Cosmologie, F-75013 Paris, France}
\affil{$^6$Department of Astronomy, University of Geneva, Switzerland}

\paperauthor{Vladislav Klochkov}{vladislavk4481@gmail.com}{0000-0003-3095-8933}{M.V. Lomonosov Moscow State University}{Department of Physics}{Moscow}{}{119991}{Russia}
\paperauthor{Ivan Katkov}{katkov.ivan@gmail.com}{0000-0002-6425-6879}{NYU Abu Dhabi}{Center for Astro, Particle, and Planetary Physics}{Abu Dhabi}{}{129188}{UAE}
\paperauthor{Igor Chilingarian}{igor.chilingarian@cfa.harvard.edu}{0000-0002-7924-3253}{Center for Astrophysics - Harvard and Smithsonian}{}{Cambridge}{}{02138}{USA}
\paperauthor{Kirill Grishin}{kirillg6@gmail.com}{0000-0003-3255-7340}{Sternberg Astronomical Institute, Lomonosov Moscow State University}{}{Moscow}{}{119234}{Russia}
\paperauthor{Anastasia Kasparova}{anastasya.kasparova@gmail.com}{0000-0002-1091-5146}{Sternberg Astronomical Institute, Lomonosov Moscow State University}{}{Moscow}{}{119234}{Russia}
\paperauthor{Vladimir Goradzhanov}{goradzhanov.vs17@physics.msu.ru}{0000-0002-2550-2520}{Sternberg Astronomical Institute, Lomonosov Moscow State University}{}{Moscow}{}{119234}{Russia}
\paperauthor{Victoria Toptun}{victoria.toptun@voxastro.org}{0000-0003-3599-3877}{Sternberg Astronomical Institute, Lomonosov Moscow State University}{}{Moscow}{}{119234}{Russia}
\paperauthor{Evgenii Rubtsov}{rubtsov602@gmail.com}{0000-0001-8427-0240}{Sternberg Astronomical Institute, Lomonosov Moscow State University}{}{Moscow}{}{119234}{Russia}
\paperauthor{Sviatoslav Borisov}{sb.borisov@voxastro.org}{0000-0002-2516-9000}{University of Geneva}{Department of Astronomy}{Geneva}{}{}{Switzerland}

\begin{abstract}
We present a set of open-source web tools for visualization of spectral and imaging data, which we use in the second Reference Catalogue of Spectral Energy Distributions of galaxies RCSEDv2 (https://rcsed2.voxastro.org/). Using modern web frameworks {\sc Quasar} and {\sc Vue.js} we developed interactive viewers to visualize spectra and SEDs of galaxies and the diagrams presenting emission line ratios determined from the analysis of their spectra (BPT diagrams). The viewers are built in {\sc Javascript} which puts a minimum load on the server side while providing full interactivity for the user. The use of modern web frameworks provides full customization making the viewers easily embeddable into web-sites of astronomical archives and databases. It also provides compatibility with popular third-party web-tools such as Aladin Lite.
  
\end{abstract}

\section{Introduction}
The rapid development of Virtual Observatory technologies raises the demand for customizable and easily embeddable data visualization tools, which can acquire data from the VO and be used as web components across different projects. With the exceedingly increasing amount of data, the need for the interactive visualisation rises accordingly. Such visualisation tools can be easily implemented with the help of modern web frameworks. In the second version of Reference Catalog of Spectral Energy Distributions of galaxies RCSEDv2 we substantially improved the data visualization solutions as compared to the first release of RCSED \citep{RCSED}. We have created tools for interactive data visualisation of 5.7 million spectra from publicly available sky surveys. 

\section{General structure of the RCSEDv2 web-site}
The RCSEDv2 structure is split in two parts, the frontend and the backend. This allows us to move the data visualisation part on the user side and eliminates the need for pre-generated templates/images/spectra. The catalog data is stored in a form of PostgreSQL tables. From there it is processed by the backend and then sent to the frontend as a JSON file to be read and then displayed.

\paragraph{The backend}
All catalog data is combined in the PostgreSQL 12.4 RDBMS. Fast cone search is available through PgSphere \citep{2004ASPC..314..225C}. The database also supports VO access interfaces via GAVO DaCHS \citep{2014A&C.....7...27D} either via Simple Spectrum Access Protocol or Table Access Protocol. The data is then processed with the use of the {\sc Django} Web Framework. User-submitted queries are processed with the {\sc Django} Custom Query. Our search query engine supports logical operators, parametric search, cone search, SIMBAD-based object name resolution and queries for the linked table columns.
\articlefigure{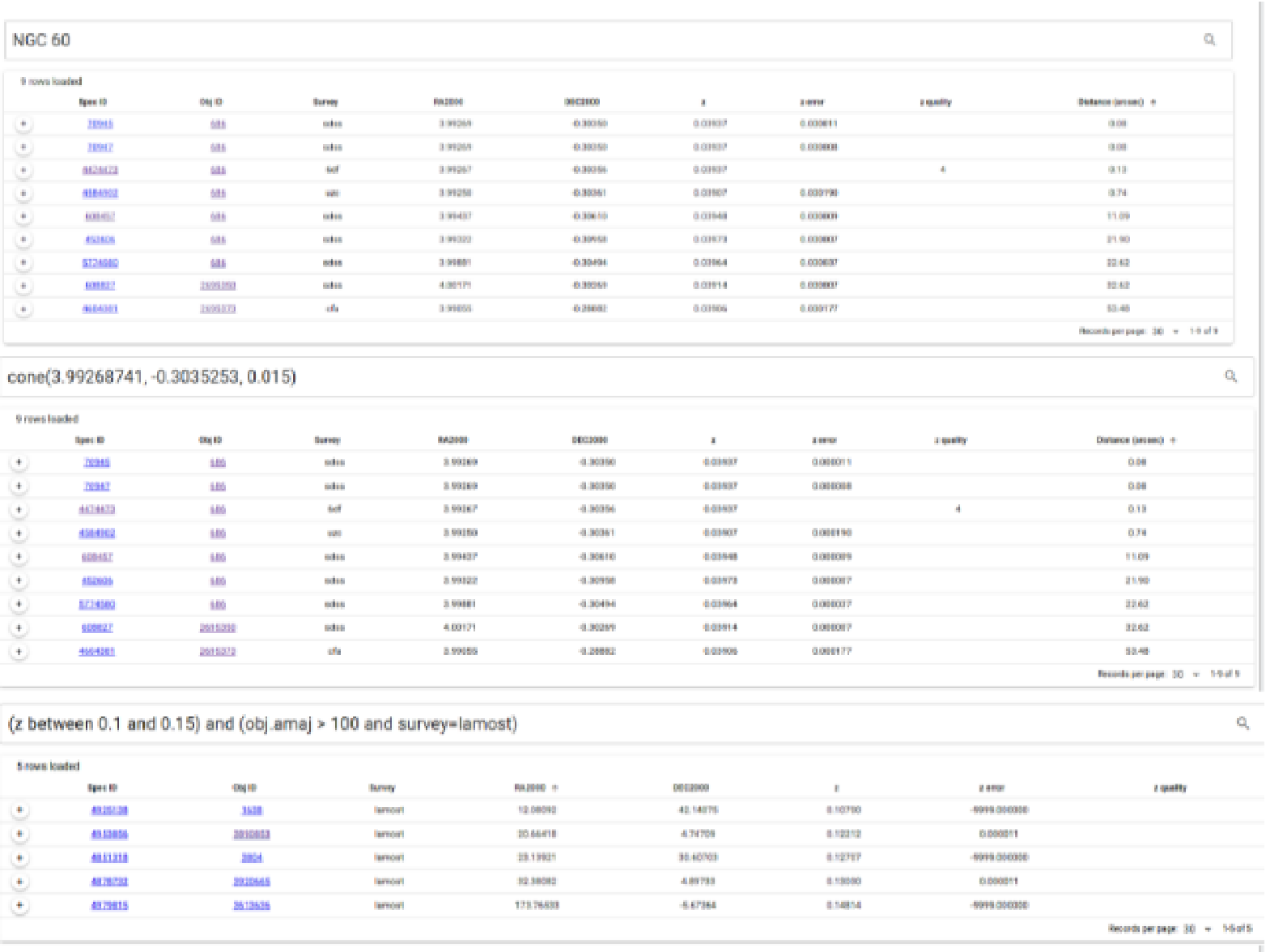}{fig1}{Examples of search queries and their results}

\paragraph{The frontend}
The frontend part of the website is implemented using the web frameworks {\sc Quasar} and {\sc Vue.js} as well as {\sc Plotly.js} for the interactive data visualisation. We implemented the tools to visualize the diagrams presenting emission line ratios determined from the analysis of their spectra (BPT diagrams, \citealp{1981PASP...93....5B}) and interactive viewers for spectra broad-band spectral energy distributions (SEDs). The spectrum viewer allows one to switch between the different modes of the full spectrum fitting, i.e. different grids of stellar population models and star formation histories as implemented in the {\sc NBursts} code \citep{CPSK07,2007MNRAS.376.1033C}, which we use for the data analysis in RCSEDv2.

\articlefigure[scale=0.45]{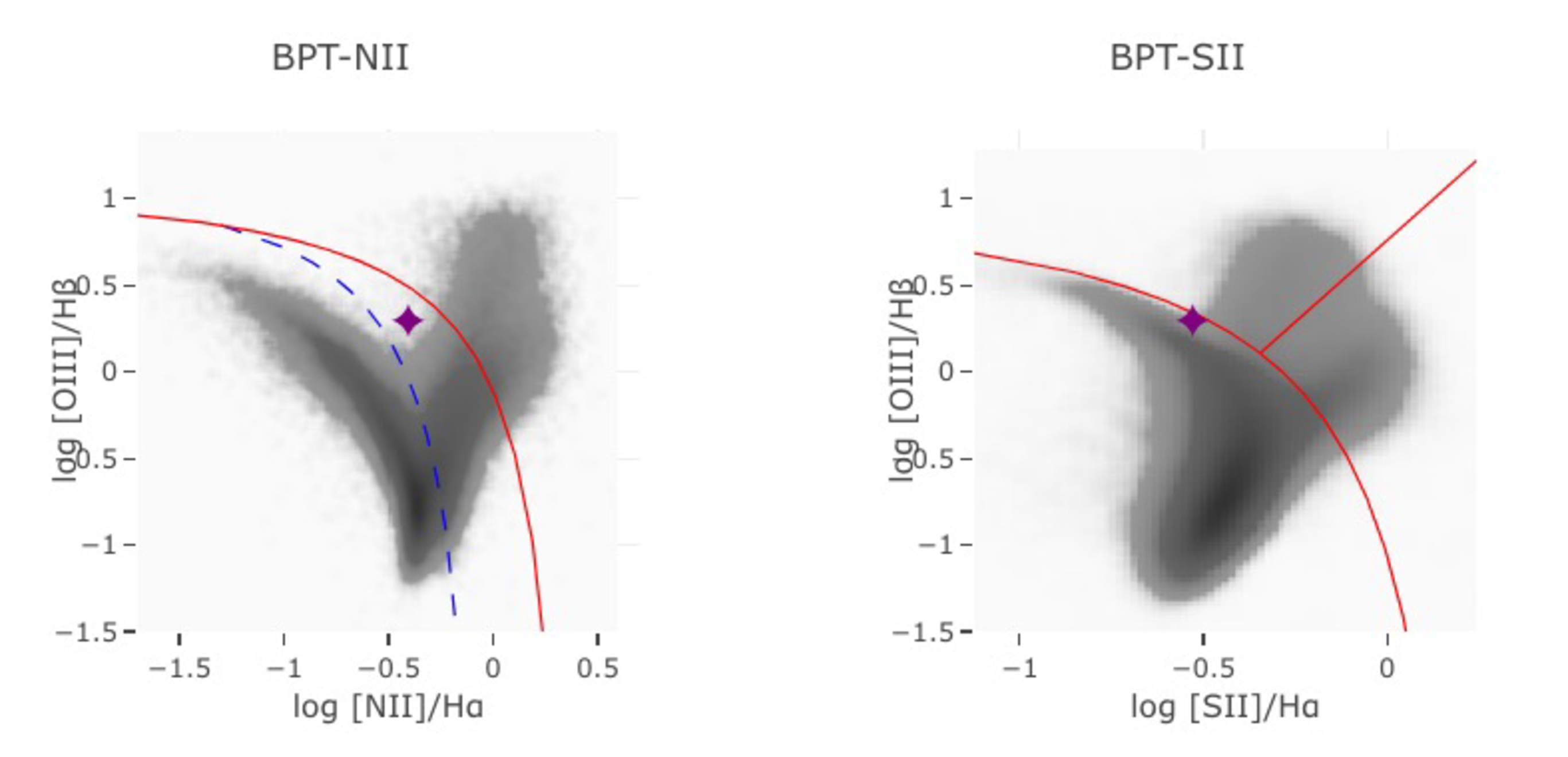}{fig2}{BPT diagrams for J110731.23+134712.8, an AGN powered by an intermediate-mass black hole in a star-forming galaxy \citep{Chilingarian+18}.}

\articlefigure{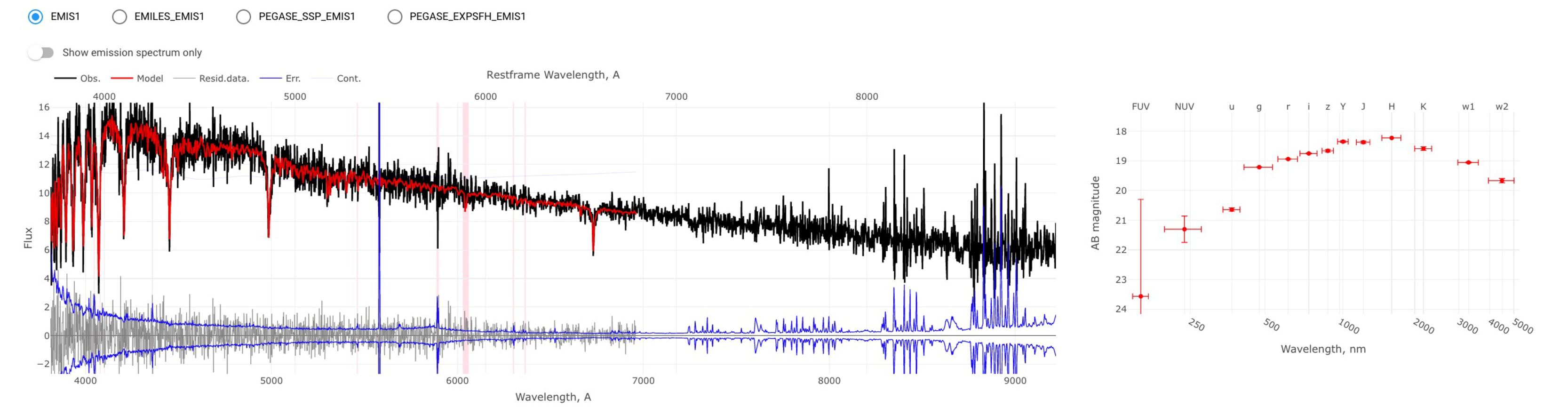}{fig3}{A spectrum with its best-fitting template and a broadband SED for GMP~4348, a dwarf post-starburst galaxy in the Coma cluster \citep{2021NatAs.tmp..208G}. }

The web frameworks we use provide compatibility with third-party web-tools such as Aladin Lite \citep{2014ASPC..485..277B}. We created an Aladin Lite-based component which allows one to switch between different imaging surveys and displays the positions of all the spectra available in RCSEDv2 as clickable markers, which forward the user to web-pages for of the corresponding objects. 

\articlefigure[scale=0.3]{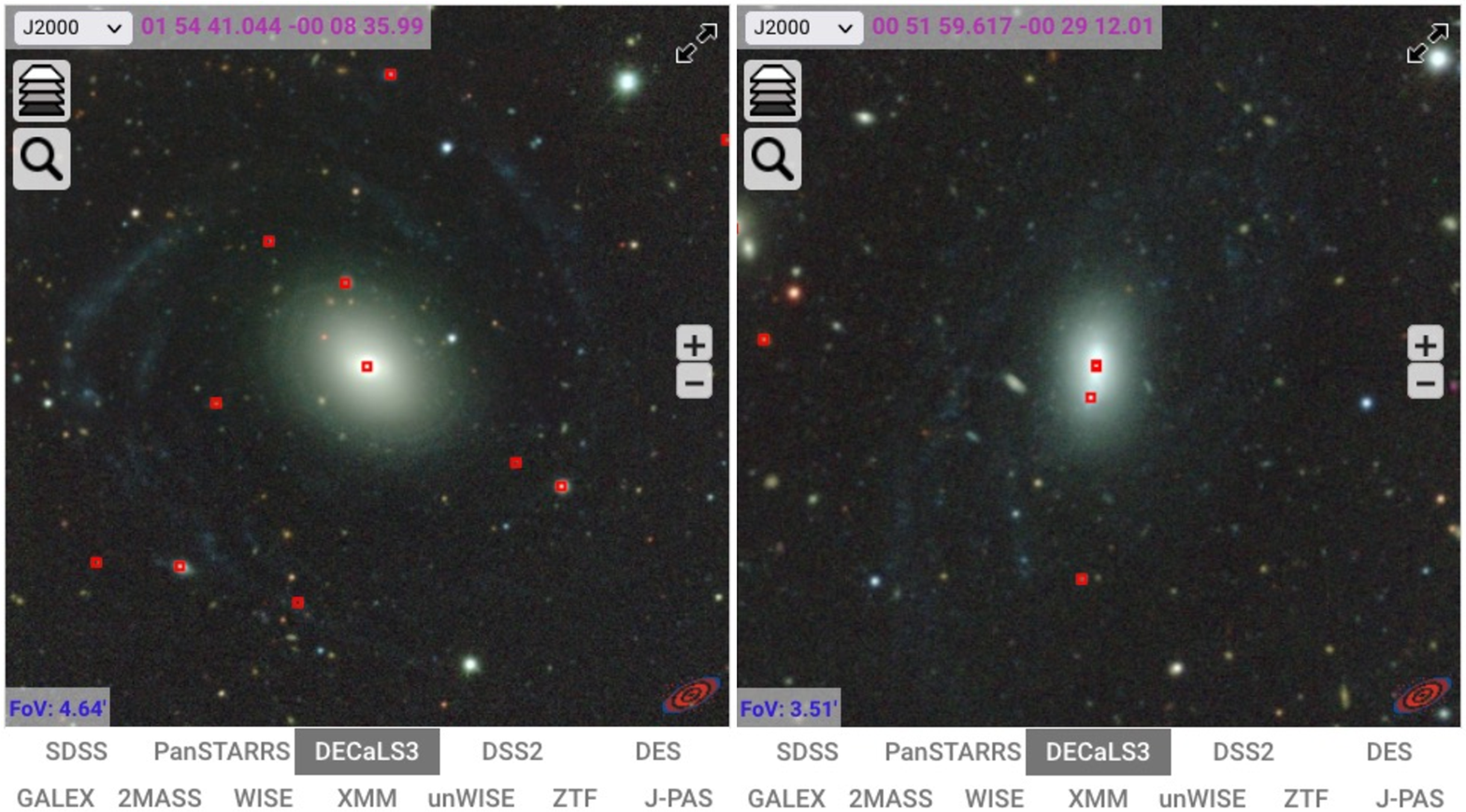}{fig4}{DECaLS images of the two low-surface brightness galaxies with the spectra available in RCSEDv2, a giant LSB UGC~1382 and a dwarf LSB Ark~18 \citep{2021MNRAS.503..830S, 2021MNRAS.504.6179E} in the embedded Aladin Lite viewer. Both galaxies have two or more spectra in different parts of their disks. In the bottom-right region of UGC~1382, there is a very rare compact elliptical (cE) satellite galaxy \citep{2015Sci...348..418C} projected onto its disk, also with the available spectrum. Such an unusual association sheds the light on the origin of giant LSBs.}

\section{Summary}
All the presented visualisation components, which we developed for RCSEDv2 can be relatively easily embedded into any web-site providing access to astronomical data for galaxies and/or stars. The flexibility and versatility of modern web frameworks allows one to develop quite complex interactive visualization solutions with a minimal investment of time and resources. Interaction between the web components such the demonstrated integration between Aladin Lite and the database query engine should definitely be considered by the developers of next generation astronomical data archives at large astronomical observatories -- this brings the user experience to the next level and, at the end, increases the science outcome of astronomical archives and databases.

\acknowledgements his project is supported by the RScF Grant 19-12-00281 the Interdisciplinary Scientific and Educational School of Moscow University ``Fundamental and Applied Space Research''.

\bibliography{O3-006}

\begin{thebibliography}{}
\expandafter\ifx\csname natexlab\endcsname\relax\def\natexlab#1{#1}\fi
\expandafter\ifx\csname url\endcsname\relax
  \def\url#1{\texttt{#1}}\fi
\expandafter\ifx\csname urlprefix\endcsname\relax\def\urlprefix{URL }\fi
\providecommand{\eprint}[2][]{\url{#2}}

\bibitem[{{Baldwin} et~al.(1981){Baldwin}, {Phillips}, \&
  {Terlevich}}]{1981PASP...93....5B}
{Baldwin}, J.~A., {Phillips}, M.~M., \& {Terlevich}, R. 1981, \pasp, 93, 5

\bibitem[{{Boch} \& {Fernique}(2014)}]{2014ASPC..485..277B}
{Boch}, T., \& {Fernique}, P. 2014, in Astronomical Data Analysis Software and
  Systems XXIII, edited by N.~{Manset}, \& P.~{Forshay}, vol. 485 of
  Astronomical Society of the Pacific Conference Series, 277

\bibitem[{{Chilingarian} et~al.(2004){Chilingarian}, {Bartunov}, {Richter}, \&
  {Sigaev}}]{2004ASPC..314..225C}
{Chilingarian}, I., {et~al.} 2004, in Astronomical Data Analysis Software and
  Systems (ADASS) XIII, edited by F.~{Ochsenbein}, M.~G. {Allen}, \&
  D.~{Egret}, vol. 314 of Astronomical Society of the Pacific Conference
  Series, 225

\bibitem[{{Chilingarian} et~al.(2007{\natexlab{a}}){Chilingarian}, {Prugniel},
  {Sil'chenko}, \& {Koleva}}]{CPSK07}
--- 2007{\natexlab{a}}, in Stellar Populations as Building Blocks of Galaxies,
  edited by A.~{Vazdekis}, \& R.~R.~{Peletier} (Cambridge, UK: Cambridge
  University Press), vol. 241 of IAU Symposium, 175. \eprint{arXiv:0709.3047}

\bibitem[{{Chilingarian} \& {Zolotukhin}(2015)}]{2015Sci...348..418C}
{Chilingarian}, I., \& {Zolotukhin}, I. 2015, Science, 348, 418.
  \eprint{1504.06990}

\bibitem[{{Chilingarian} et~al.(2018){Chilingarian}, {Katkov}, {Zolotukhin},
  {Grishin}, {Beletsky}, {Boutsia}, \& {Osip}}]{Chilingarian+18}
{Chilingarian}, I.~V., {et~al.} 2018, \apj, 863, 1. \eprint{1805.01467}

\bibitem[{{Chilingarian} et~al.(2007{\natexlab{b}}){Chilingarian}, {Prugniel},
  {Sil'Chenko}, \& {Afanasiev}}]{2007MNRAS.376.1033C}
--- 2007{\natexlab{b}}, \mnras, 376, 1033. \eprint{astro-ph/0701842}

\bibitem[{{Chilingarian} et~al.(2017){Chilingarian}, {Zolotukhin}, {Katkov},
  {Melchior}, {Rubtsov}, \& {Grishin}}]{RCSED}
--- 2017, \apjs, 228, 14. \eprint{1612.02047}

\bibitem[{{Demleitner} et~al.(2014){Demleitner}, {Neves}, {Rothmaier}, \&
  {Wambsganss}}]{2014A&C.....7...27D}
{Demleitner}, M., {et~al.} 2014, Astronomy and Computing, 7, 27.
  \eprint{1408.5733}

\bibitem[{{Egorova} et~al.(2021){Egorova}, {Egorov}, {Moiseev}, {Saburova},
  {Grishin}, \& {Chilingarian}}]{2021MNRAS.504.6179E}
{Egorova}, E.~S., {et~al.} 2021, \mnras, 504, 6179. \eprint{2103.00211}

\bibitem[{{Grishin} et~al.(2021){Grishin}, {Chilingarian}, {Afanasiev},
  {Fabricant}, {Katkov}, {Moran}, \& {Yagi}}]{2021NatAs.tmp..208G}
{Grishin}, K.~A., {et~al.} 2021, Nature Astronomy. \eprint{2111.01140}

\bibitem[{{Saburova} et~al.(2021){Saburova}, {Chilingarian}, {Kasparova},
  {Sil'chenko}, {Grishin}, {Katkov}, \& {Uklein}}]{2021MNRAS.503..830S}
{Saburova}, A.~S., {et~al.} 2021, \mnras, 503, 830. \eprint{2011.01238}

\end{thebibliography}


\end{document}